International Conference On Applied Economics (ICOAE) 2012

# Does Non-Farm Income Improve The Poverty and Income Inequality Among Agricultural Household In Rural Kedah?

Siti Hadijah Che Mat[a]*, Ahmad Zafarullah Abdul Jalil[b], Mukaramah Harun[c]

*a, b, c College of Business, Universiti Utara Malaysia*

**Abstract**

This paper used a primary data collected through a surveys among farmers in rural Kedah to examine the effect of non farm income on poverty and income inequality. This paper employed two method, for the first objective which is to examine the impact of non farm income to poverty, we used poverty decomposition techniques- Foster, Greer and Thorbecke (FGT) as has been done by Adams (2004). For the second objective, which is to examine the impact of non farm income to income inequality, we used Gini decomposition techniques. Our result indicate that non farm income can improve the level of poverty or non farm income sources contributed towards poverty reduction among agricultural household. All of the poverty measures show that the inclusion of non-farm income into the agricultural household income reduce the level, depth and severity of poverty. But on the other hand, non farm income increased income inequality among agricultural household in Kedah. As expected agricultural income is the main source of income for rural people in the study area. The policy implication of this study is to encourage non-farm income activities among agricultural households as this would raise their income and hence, reduce poverty among them. However, it should be focused on value-added activities, especially on the lower income group.

**Keywords:** non-farm income, poverty, inequality, Malaysia

* Corresponding author. Tel.:604-9283544; fax:604-9286346
*E-mail address:hadijah@uum.edu.my.*

1. Introduction

Non-farm income is important for poverty reduction (de Janvry and Sadoulet, 2001) and for improving household welfare (Reardon et al., 1992; Reardon et al., 1998; Barrett et al., 2000;Canagarajah et al., 2001). Haggblade et al. (2005), for instance, report that non-farm income contributed 30 - 45 percent of rural household income across the developing world. Based on a review of a number of studies using rural household surveys conducted between the mid 1970s and the late 1990s, Reardon et al. (1998) finds that non-farm income as a share of total household income averaged 42 percent for Africa, 32 percent for Asia and 40 percent for Latin America. In Ghana, non-farm employment is an equally important source of income for rural households. On the other hand the empirical evidence on the effect of non-farm income on rural income inequality is mixed. Canagarajah et al. (2001) observes that this result may be due to the heterogeneity of the non-farm sector.

Malaysia has made remarkable progress in reducing poverty since the introduction of a New Economic Policy 1970 with the headcount poverty rate declining from 37.7 percent in 1976 to 3.6 percent in 2007. But in terms of income distribution as shown by the Gini coefficient found that in almost 40 years, the improvements shown in the income distribution is very small. The Gini coefficient decreased slightly about 0.072 from 0.513 in 2007 to 0.441 in 2007. The question now, is this imbalance in income distribution is related to the non- farm activities. Specifically in this paper, we are trying to find the percentage of farmer having the source of non farm income. And to what extent the income received from non farm activities can increase farmer's income and thus reduce poverty and to what extent the non farm income effect the income inequality.

2. Literature review

The empirical evidence on the effect of non-farm income on rural income inequality shows mixed results. Studies such as Reardon and Taylor (1996), Reardon et al. (1998), Adams (2001), Elbers and Lanjouw (2001) and Woldehanna (2002) find that non-farm income increases inequality because non-farm income is unequally distributed in favor of the rich. Lanjouw (2000), found in the state of Ecuador, non-farm sector contributed 40% of rural incomes. Nearly 40% of men and 50% of the women involved in this activity and also income from non-farm employment is associated positively with the level of education and infrastructure. De Janvry (1981), wages from a part time job in rural areas often complement the inadequacy of agricultural production to ensure the needs of household consumption will continue to be enjoyed. De Janvry et al. (2005), studies in China, involving 7041 households with agricultural and non agricultural income showed, 72% of rural households have non-farm income. Non-farm income is not only able to absorb surplus labor in rural areas, but more importantly what it can improve is the quality of life in rural areas. It can be concluded that non-farm income can be considered as a potential successor to the agricultural income. On the other hand, Adams (1994), Lanjouw (1998) and Zhu and Luo (2006) find that non-farm income decreases rural income inequality. Reardon et al. (2000) observes that the assertion that non-farm income reduces income inequality is premised on three empirical assumptions: 1) that non-farm income is large enough to influence rural income distribution, 2) that non-farm income is unequally distributed, and 3) that this unequally distributed non-farm income favors the poor.

Non-farm income also plays an important role and exhibits an increasing share in agricultural household income (De Janvry et.al, 2005; FAO, 1988). Thus, the non-farm (or off-farm) employment has been generally recognised to have the potential in raising agricultural household income, and therefore reducing rural poverty (FAO, 1998; Arif, Nazli and Haq, 2000; Lanjouw and Murgai, 2008; Foster and Rosenzweig, 2004). Non-farm income gradually became an importance source of income for rural households, and served as an engine of growth for rural areas. Adams (2001) on his study at Egypt and Jordan, find that non-farm income has a greater impact on poverty and inequality. The poor receive almost 60 percent of their income from non-farm sources in rural Egypt, while in rural Jordan they receive less than 20 percent. A study done by Roslan and Siti Hadijah (2011), found that farmers that participate in non-farm activities, has a clearly shorter their average time to exit poverty than those who did not participate in non-farm activities.

## 3. Data and Methodology

The state of Kedah was chosen because there are many people involved as a farmers compared to the states in Peninsular Malaysia. Kedah also have highest poverty rate in Peninsular Malaysia. For example in 2004, Kedah recorded a high overall poverty level of 7% (27.300 households). Poverty is above the overall poverty rate of 5.7%. (Malaysia, 2006). Agriculture is one of the main economic sector in the northern states, particularly in Kedah. According to the UNDP report of 2004 and the 9th Malaysia Plan (2006), the workforce engaged in agriculture in the Northern Territory Malaysia by state is 21.7% of the workforce in Perlis, Kedah 19%, 18% Perak and 1.4% on the Island Penang.

The data used in this study is primary data which is gathered through a survey carried out on 384 agricultural households in the state of Kedah, Malaysia. The survey is conducted between the month of April and December 2008. A face to face interview were carried out with the respondents, where they were chosen through a stratified random sampling. Six of the eleven districts in Kedah were chosen in this study. These are Kubang Pasu, Sik, Kota Star, Baling, Kulim dan Pulau Langkawi. For each district, the respondent is divided further according to the local economic characteristics (economic structure of the local economy), to see the importance and strength of the existence of industry and agriculture to the farmers to engage in non-farm activities.

In this paper we divide the total income receive by the household in to three separate item, one is for agricultural income, second is income from non-farm activities and the third one is unearned income (transfer payment). Non-farm income is the income receive by agricultural household in remunerative work away from their plot of agricultural land (FAO, 1988). The non-farm job undertaken by the farmer could be permanent or casual in nature, covering both the secondary and tertiary sector of employment (Salter, 1991). Besides, to disaggregate the poor from the non-poor, poverty line income is used. The official gross poverty line income for the state of Kedah in 2009 is RM700[*]. Thus, in this study, a farmer with a household income that is equal or more than RM700 is considered non-poor, while those with household income that is less than RM700 is categorised as poor.

For the first objective, this study used FGT index. With the modification of this index, according to Foster-Greer-Thorbecke (1984) it can be used to observe the effects of non-agricultural income on poverty. As Huppi and Ravallion (1991) used this index to see the effect of sources of income on poverty in Indonesia. While Reardon and Taylor (1996) make the separation on the FGT poverty, according to sources of income. By following Adams (2004), this study used Foster-Greere-Thorbecke (hereafter FGT) poverty index (1984) to measure the impact of non-farm income on poverty. The FGT poverty measure is defined as:

$$P_\alpha = \frac{1}{n} \sum_{i=1}^{m} \left( \frac{z - y_i}{z} \right)^\alpha$$

---

[*] e-SINAR.Kedah.gov.my

Where *n* is the whole sample used in this study, *m* is the total number of households living under the poverty line, $y_i$ is represents the income of the poor household from *i* to *m* which arrange in increasingly order, *z* is the poverty line income and *α* is a poverty aversion parameter. The three parameter (depending on three values of α) of the poverty index used to estimate the impact of changes in non-farm income and unearned income on poverty are: the headcount ratio index (*α*= 0) which is measure the share of population living below the poverty line; the poverty gap index (*α* = 1) to measures the depth of poverty, that is the amount by which an average poor family is below the poverty line. The poverty gap squared index (*α* = 2) to measures the severity of poverty and, unlike the other two measures, is sensitive to changes in the distribution of income among the poor (Adams and Page 2005).

In this study, income of the household are divided into four groups, as done by Adams (2004). The first group is the farmer who have only agricultural income. The second group is income from agricultural income plus non-farm income, the third group is income receive by farmer from agricultural activities and unearned income and the last group is total income receive by farmer.

For the second objective, i.e to see the impact of non farm income on income inequality, this study used decomposition of Gini coefficient. Any reliable measure of inequality must meet five basic properties, namely, (1) Pigou-Dalton transfer sensitivity; (2) symmetry or anonymity; (3) mean independence; (4) population homogeneity; and (5) decomposability, (Litchfield, 1999). A measure of income inequality that meets all the above properties is the Gini coefficient. Decomposability is the property that requires overall inequality to be partitioned into its constituent parts, either over sub-populations or sources. That is, an inequality measure can be regarded as source decomposable if total inequality can be broken down into a weighted sum of inequality by various income components. The income equalizing or dis-equalizing effect of non-farm income can therefore be ascertained by decomposing the Gini coefficient of total income into its component parts.

Following Lerman and Yitzhaki (1985), the Gini coefficient for total income inequality, G, can be represented as:

$$G = \sum_{k=1}^{K} R_k G_k S_k$$

Where $S_k$ represents the share of component k in total income, $G_k$ is the source Gini, corresponding to the distribution of income from source k, and $R_k$ is the Gini correlation of income from source k with the distribution of total income.

## 4. Findings

Table 1 report three different poverty index by using FGT index. All of the poverty measures show that the inclusion of non-farm income and un earn income into the agricultural household income reduce the level, depth and severity of poverty in Kedah. However the size of the poverty reduction depends very much on how poverty is measured. According poverty headcount measure, including non farm income to income agricultural household reduces the level of poverty by 42.94 percent and inducing un earn income in agricultural household income reduces the level of poverty by 14.72 percent. However, poverty is reduced much more when it measured by the depth and severity of poverty, such as the poverty gap and squared poverty gap. For example, the squared poverty gap measure shows that including non farm income and un earn income in household agricultural income reduce poverty by 55.71 or 23.35 percent, respectively.

**Table 1**
**FGT index: The Effect of Non-Farm Income on poverty in Kedah.**

*Source:* Researcher estimates on 381 household

With respect to poverty, Table 1 shows that non farm income has a slightly greater impact on poverty than un earn income. For instance, all three poverty measures shows that the extent of poverty reduction is

| PLI/ α | | Farm income only @ Farminc (1) | Farm income and Non farm income only @ Farminc + NFi (2) | Farm income and un earn income only @ Farminc + Transinc (3) | Total house hold income @ Farminc + NFi+ Transinc (4) | % Change (Farminc vs. ada NFi) [(2)-(1)]/(1)* 100 (5) | % Change (Farminc vs. Transinc) [(3)-(1)]/(1)* 100 (6) | % Change (Farminc vs NFi dan Transinc) [(4)-(1)]/(1)* 100 (7) |
|---|---|---|---|---|---|---|---|---|
| 700 | 0 | 0.4094 | 0.2336 | 0.3491 | 0.1811 | -42.9409 | -14.7289 | -55.7645 |
|  | 1 | 0.1805 | 0.0876 | 0.1456 | 0.0663 | -51.4681 | -19.3352 | -63.2687 |
|  | 2 | 0.1032 | 0.0457 | 0.0791 | 0.0344 | -55.7171 | -23.3527 | -66.6667 |

greater when non farm income are included in agricultural household income, as opposed to when un earn income are included. When both non farm income and un earn income are included in agricultural household income, the effect on the poverty reduction is high. The reduction in head count ratio, depth of poverty and squared poverty gap index are 55.76 percent, 63.27 percent and 66.67 percent respectively.

Table 2, shows the results of Gini decomposition analysis. The column (1) of (Sk) shows the contribution of particular income sources to overall income. As expected agricultural income is the main source of income for rural people in the study area. Agricultural income contributes more than 60 percent of the total income meanwhile non-farm income is 32 percent.

**Table 2**
**Gini Decomposition by Income Source**

| Income Source | Share in Total income ($S_k$) (1) | Gini Coefficient ($G_k$) (2) | Gini Correlation ($R_k$) (3) | Relatif Contribution ($S_k*R_k*G_k/G$) @share in total income inequality (4) | Marginal effect / % change in Gini from a % change in income source (5) |
|---|---|---|---|---|---|
| *Farminc* | 0.6221 | 0.4804 | 0.8633 | 0.6188(61.88) | -0.0033 |
| *Ofarminc* | 0.3235 | 0.6475 | 0.7203 | 0.3619 (36.19) | 0.0385 |
| *Transinc* | 0.0544 | 0.7765 | 0.1897 | 0.0192 (1.92) | -0.0352 |
| *Tot.Income* | 1.000 | 0.4169 | - | 1.000 (100.00) | - |

*Source*: Researcher estimates on 381 household.

The column (2) is Gini coefficient (Gk), which showed equity in income distribution or inequality for each source of income. The first line is agricultural income (Farminc), shows the distribution of income from agricultural sources received by households without the presence of non-farm income (Ofarminc) and non-employment income (Transinc). By comparing the figures (Gk) for each source to the total Gk, we get the effect of that particular income (increase / decrease) to total inequality. In this study, the overall income inequality dropped when the non-farm income and un earn income is taken into consideration together with agricultural income. For agricultural income, the Gini coefficient dropped by 6:35 per cent - from 0.4804 to 0.4169.

The results show, non-farm income (Ofarminc) contribute about 32.35% to total income, but what is worrying is source of income is not distributed equally as shown by Gk (0.6475) is much more higher than the value of Gk (0.4804) farm income. Although the distribution of income by non-farm sources of income shows a high inequality, but relatively non-farm income is only accounted for 36.19 % to overall income inequality compared to the percentage contribution of agricultural income is much larger (61.88 percent) to total inequality.

The effect on income inequality when there is a small change in a source of income can be seen in column (5). This study showed that a 1 percent increase in agricultural income sources (farminc), assuming other sources of income unchanged, will reduce the overall inequality by 0.3 percent. Similarly, the unearned income (Transinc), a 1 percent increase in un earn sources of income will be reduced by 3.5 percent overall inequality. In contrast to non-farm income (Ofarminc) which 1% increase in non-farm sources of income, assuming other source of income unchanged, will result in an increase of 3.9 percent overall inequality. This result is consistent with findings by Adams (2001) who found that non-farm income in Jordan. The uneven distribution of land may lead to non-farm income inequality. Among those who have vast tracts of land (over 3.5 recesses), which is 31 per cent, may be able to create non-farm income is higher than those who have land less than 3.5 recesses (69 percent). Adams (2001) found that in Egypt, the factor 'land', which is distributed unevenly on the negative and significant in determining the non-farm income.

## 5. Conclusions

This paper used a survey on agricultural household in Kedah to examine the effect of non farm income on poverty and income inequality. It is indicate that non-farm income sources contributed about one-third of the total agricultural household income. Besides, decomposition exercises carried out on poverty index shows that non-farm income sources contributed significantly towards poverty reduction among agricultural household. Meanwhile, the decomposition of the Gini coefficient analysis in this study found non-farm income sources contributed to increase income inequality or dis-equalising factor among the agricultural households. While farm and unearned income are income equalizers. The policy implication of this study is that, non-farm income activities should be encouraged among agricultural households as this would raise their income and hence, reduce poverty among them. However, it should be focused on value-added activities, especially on the lower income group. The findings of this study also suggest that a balanced development approach should not only focus on rural-urban divide, but also within the rural areas itself.

Siti Hadijah Che Mat et al. / Procedia Economics and Finance 1 (2012) 269 – 275	275